\documentclass[10pt,aps,prl,twocolumn,groupedaddress,showkeys]{revtex4-1}
\usepackage{amsmath}
\usepackage{graphicx}
\DeclareGraphicsExtensions{.pdf,.eps,.png,.jpg,.mps}
\usepackage{epstopdf}
\usepackage{threeparttable}
\usepackage{color}
\begin{document}
\title{Non-equilibrium Band Broadening, Gap Renormalization and Band Inversion in Black Phosphorus}
\author{H. Hedayat$^1$, A. Ceraso$^{1,2}$, G. Soavi$^{3,4}$,S. Akhavan$^3$,A. Cadore$^3$,C. Dallera$^2$, G. Cerullo$^{1,2}$,A.C. Ferrari$^3$, E. Carpene$^1$}
\affiliation{$^1$Institute for Photonics and Nanotechnologies IFN-CNR, piazza Leonardo da Vinci 32, 20133 Milano, Italy}
\affiliation{$^2$Dipartimento di Fisica, Politecnico di Milano, 20133 Milano, Italy}
\affiliation{$^3$Cambridge Graphene Centre, University of Cambridge, Cambridge CB3 0FA, UK}
\affiliation{$^4$Institute for Solid State Physics, Abbe Center of Photonics, Friedrich-Schiller University Jena, 07743 Jena, Germany}
\begin{abstract}
Black phosphorous (BP) is a layered semiconductor with high carrier mobility, anisotropic optical response and wide bandgap tunability. In view of its application in optoelectronic devices, understanding transient photo-induced effects is crucial. Here, we investigate by time- and angle-resolved photoemission spectroscopy BP in its pristine state and in the presence of Stark splitting, chemically induced by Cs ad-sorption. We show that photo-injected carriers trigger bandgap renormalization and a concurrent valence band flattening caused by Pauli blocking. In the biased sample, photo-excitation leads to a long-lived (ns) surface photovoltage of few hundreds mV that counterbalances the Cs-induced surface band bending. This allows us to disentangle bulk from surface electronic states and to clarify the mechanism underlying the band inversion observed in bulk samples.
\end{abstract}
\maketitle
Black phosphorus (BP) is a layered semiconductor with outstanding physical properties such as high carrier mobility (up to$\sim10^4$cm$^2$V$^{-1}$s$^{-1}$ in the monolayer(1L))\cite{qiao}, large electronic/optical anisotropies (reflectance and DC conductance can vary by a factor$\sim2-4$ with in-plane orientation)\cite{xia,xin} and excellent mechanical properties (1L-BP can sustain tensile strain up to$\sim$30\%)\cite{wei}. Its direct bandgap depends on the number of layers\cite{rudenko,tran}, ranging from$\sim0.4$eV in bulk\cite{morita} to $\sim$2eV in 1L-BP, phosphorene\cite{tran}, and is  sensitive to pressure\cite{press1,press2}, electric field\cite{liu2,liu3,dolui,deng} and in-plane strain\cite{li,rodin}. Refs.\citenum{kim,ehlen,swkim} demonstrated that surface doping by alkali atoms allows to engineer the gap of BP, leading to surface band inversion at a critical dopant concentration$\sim0.4$1L ($\sim9\times10^{13}$cm$^{-2}$)\cite{kim}. This gap tunability is attributed to the so-called giant Stark effect\cite{gse1,gse2}, i.e. an electric field-induced shift of electronic states, named "giant" as it can lead to gap closure, as confirmed by modelling\cite{liu2,swkim,ehlen,rudenko} and photoemission experiments\cite{ehlen,hof}, and a pronounced surface depletion at the valence band (VB) and surface confinement of the conduction band (CB).

In light of the promise of BP for opto-electronic applications\cite{xia,yuan,chen}, it is important to understand its ultrafast non-equilibrium response. Time- and angle-resolved photoemission spectroscopy (TARPES), exploiting a pump-probe scheme, can track the dynamics of the electronic structure after an ultrashort (tens of fs) optical stimulus\cite{book}. To date, only a few TARPES studies have been performed on BP.
The role of photo-induced band broadening and ionized surface impurities on carrier dynamics was studied in Ref.\citenum{perfetti}, suggesting the absence of bandgap renormalization (BGR)\cite{bgr2,bgr3}. On the other hand, the VB shift triggerd by optical excitation was attributed to BGR and corroborated by ab-initio calculations in Ref.\citenum{grioni}, in agreement with resonant transient absorption measurements\cite{miao}.
Although extensive theoretical work was done on 1L- and few-layers (FL) BP\cite{rudenko,tran,liu2,liu3,dolui,li,swkim}, ARPES measurements were only performed on bulk crystals cleaved {\it in situ} with no control on sample thickness\cite{kim,ehlen,perfetti,grioni,kimura}. The depth sensitivity of photoemission critically depends on the photon energy $h\nu$ (it can range from 1L at $h\nu\sim100$eV to several nm at $h\nu\sim6$eV)\cite{inelastic}. ARPES cannot ignore the presence of underlying bulk states, and the comparison with 1L- or FL-BP theoretical predictions might be misleading.

Here we use TARPES with 6eV probe photons to investigate the ultrafast response of photo-excited bulk BP, with and without a vertical electric bias, resulting from Cs adsorption, to trigger Stark splitting. Photo-injected electron ($e$)$-$hole ($h$) pairs thermalize within a few ps and induce VB broadening. By comparing the dynamics of pristine and alkali-adsorbed samples, we assign the broadening to carriers drift-diffusion processes, and not to the previously claimed Stark-related effects\cite{perfetti}. Our analysis also reveals$\sim50$meV BGR entwined to a transient VB flattening that arises from band filling. In biased BP, pump photons trigger SPV of a few hundreds mV, in agreement with TARPES measurements at fixed pump-probe delay\cite{perfettiSPV}. We demonstrate that SPV can partially or totally compensate Cs-induced band bending. The bulk sensitivity of our experiment allows us to disentangle bulk from surface electronic states and to establish that band inversion is a crossing between {\it surface} CB and {\it bulk} VB. Thus, the semiconductor-to-semimetal transition upon alkali surface doping can be argued only in 1L- or FL-BP\cite{swkim}, where bulk contributions are absent.
\section{Results and discussion}
\begin{figure*}
\centerline{\includegraphics[width=180mm]{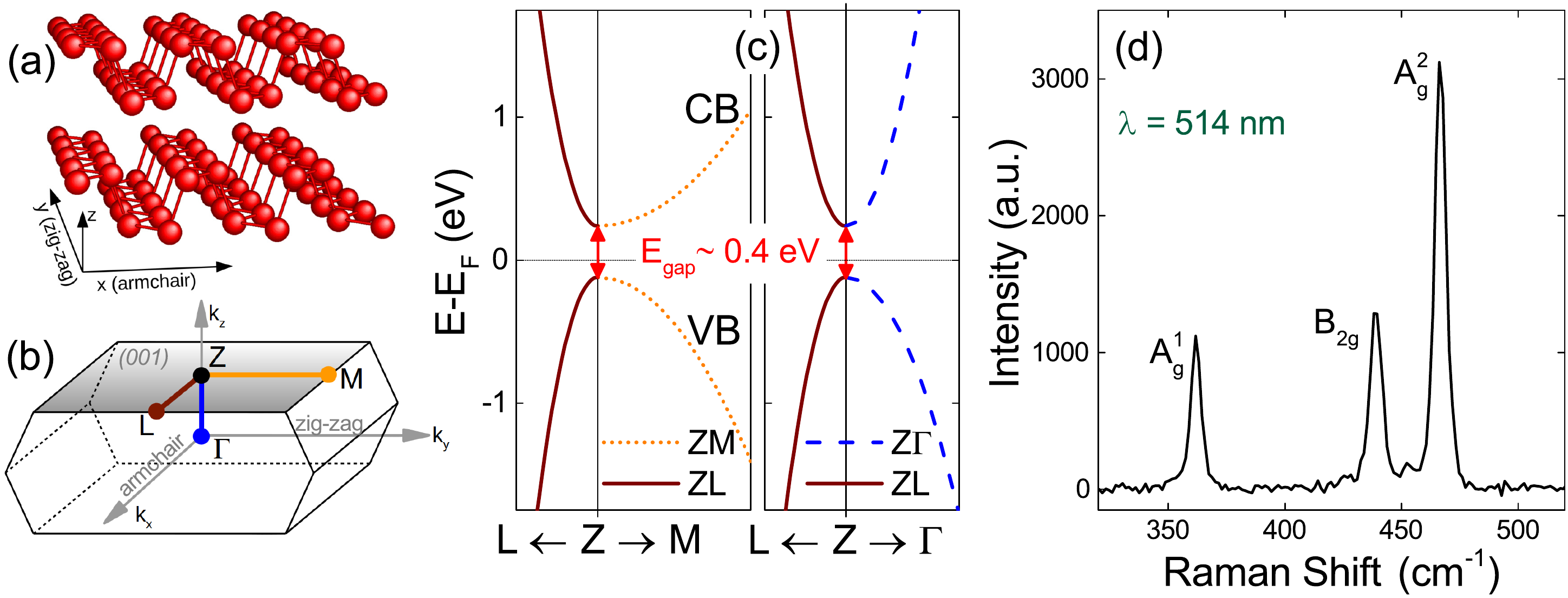}}
\caption{(a) BP crystal structure. Each phosphorene layer, normal to the $z$-axis, is buckled along the $x$-axis (armchair direction). (b) First BZ of bulk BP. The direct bandgap is located at $Z$. (c) Schematic electronic band structure around $Z$. The band dispersions along the three main axes are estimated from the effective masses of Ref.\citenum{morita}. (d) RT 514.5nm Raman spectrum of bulk BP}
\label{f1}
\end{figure*}
BP crystallizes in the orthorhombic structure\cite{morita,endo}, where $sp^3$ orbital hybridization leads to buckled layers normal to the $z$-axis, Fig.\ref{f1}a. In the bulk, the direct bandgap is located at the $Z$ point of the first Brillouin Zone (BZ)\cite{morita}, Fig.\ref{f1}b, and is strongly anisotropic, Fig.\ref{f1}c. The $e/h$ effective masses range from$\sim0.08\times m_0$ along the $x$-axis (armchair direction, where $m_0$ is the free electron mass)\cite{morita} to $\sim1\times m_0$ along the $y$-axis (zigzag direction)\cite{morita}. Here, we consider the armchair ($x$) direction, where band dispersion is largest (smallest effective mass), but our conclusions on band dynamics are unaffected by the in-plane crystallographic orientation because we probe the region surrounding the $Z$ symmetry point.

We use bulk BP from HQ graphene. Samples are characterized by Raman spectroscopy\cite{acf,acf2,ramankim} utilizing a Renishaw InVia spectrometer equipped with a 50$\times$ objective (numerical aperture NA$=0.75$) at 514nm. The laser power is kept below 100$\mu$W to avoid any possible damage\cite{lu}. Before measuring, the surface oxidized layer is pulled off via micromechanical cleavage by tape\cite{kostyaPNAS,favron,gomez}. The Raman spectrum of BP exhibits three major peaks, as shown in Fig.\ref{f1}d. They correspond to in-plane B$_{2g}$ and A$^{2}_{g}$ and out-of-plane A$^{1}_{g}$ vibrational modes\cite{gomez,rib}. Our BP flakes show A$^{1}_{g}$, B$_{2g}$ and A$^{2}_{g}\sim361.9$, 439.1, and 466.7 cm$^{-1}$, respectively, in good agreement with bulk BP literature\cite{favron,gomez,rib,sugai}.

TARPES is performed with the setup described in Ref.\citenum{rsi}. The laser source, based on a Yb system (Pharos, Light Conversion) and a non-collinear optical parametric amplifier, provides ultrashort pump ($h\nu=1.82$eV, pulse duration 30fs) and probe ($h\nu=6$eV, pulse duration 70fs) pulses at 80kHz repetition rate. $p$-polarized pump and probe beams, focused respectively on spots$\sim135$ and $\sim60\mu$m in diameter, impinge at $45^{\circ}$ on the sample cleaved in-situ under ultrahigh vacuum conditions ($\sim10^{-10}$mbar). Photo-emitted $e$ are detected by a hemispherical analyzer (Phoibos 100, Specs), with a combined energy resolution$\sim45$meV (as estimated from the low energy cut-off of the spectra) and angular acceptance$\pm15^{\circ}$. All data are recorded at room temperature (RT). The in-plane crystallographic orientation of the sample is determined by Low Energy Electron Diffraction and cross-checked by exploiting the in-plane anisotropy of the VB dispersion seen by ARPES.

Cs deposition is performed with a SAES Getters dispenser inside the preparation chamber adjacent the photoemission one\cite{rsi}. The current through the dispenser ($\sim5$A) is adjusted to keep$\sim10^{-9}$mbar during deposition (with starting pressure$\sim10^{-10}$mbar). The doping dose, proportional to the deposition time, is estimated comparing the resulting bandgap with the data in Ref.\citenum{kim}. This indicates that 45s correspond to the critical dose$\sim0.35$1L Cs, that closes the bandgap.
\begin{figure*}
\centerline{\includegraphics[width=180mm]{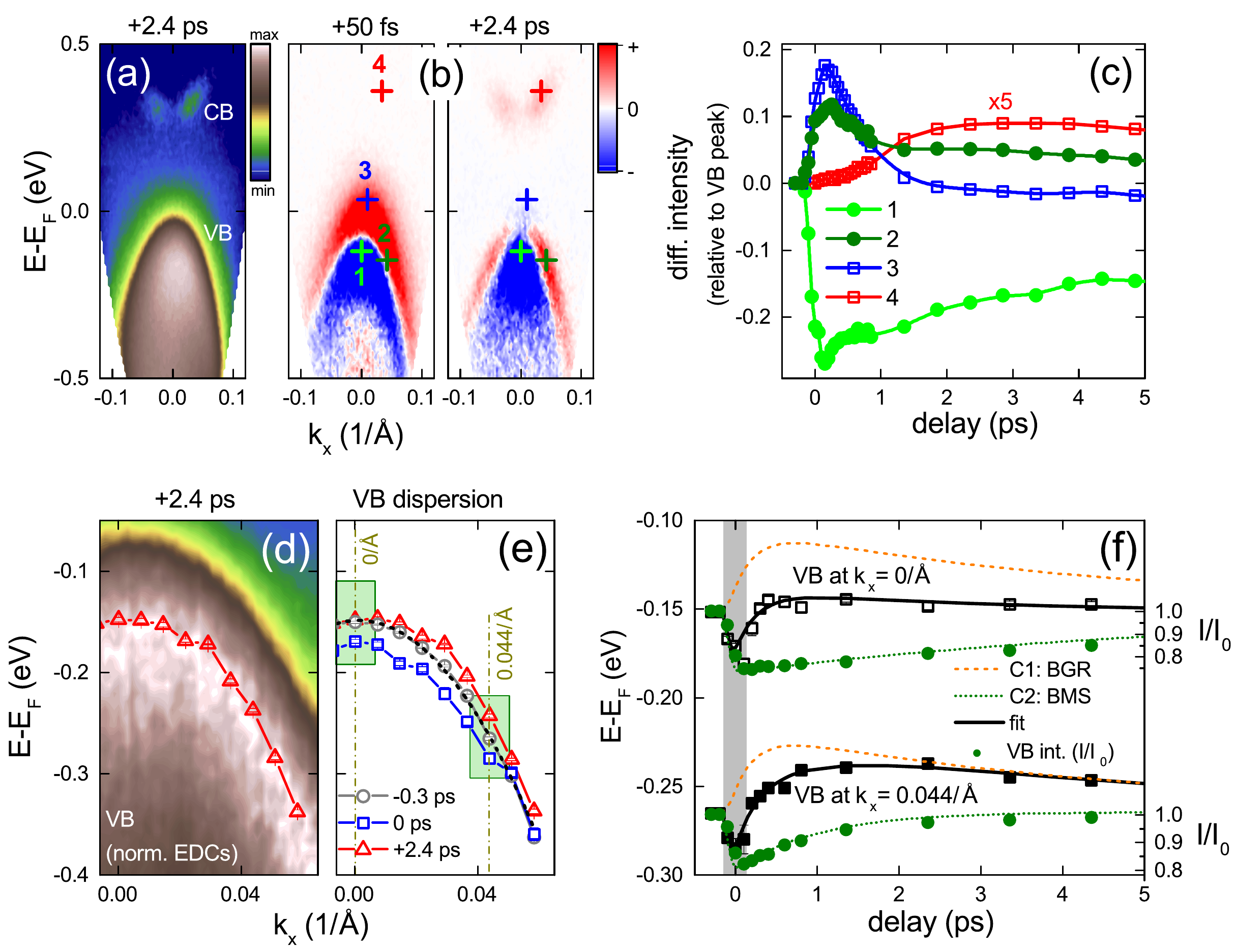}}
\caption{(a) ARPES map of BP along the $x$-axis, Fig.\ref{f1}, at pump-probe delay +2.4ps, where also CB is populated. (b) Differential spectra at two selected delays: red represents enhancement, blue is depletion. The crosses labeled 1 to 4 mark spectral features whose dynamics is shown in panel (c) (solid lines are guides to the eye). (d) Close-up of VB maps at +2.4ps delay. EDCs are normalized to their maxima. The symbols represent the fitted band dispersions (see Methods). (e) Comparison of VB dispersions at -0.3ps (circles), 0ps (squares) and +2.4ps (triangles) delays. The black dashed line is a parabolic fit. (f) VB dynamics at $k_x=0$\AA$^{-1}$ (open squares) and $k_x=0.044$\AA$^{-1}$ (solid squares). Lines are model results from a phenomenological fit (see text), while solid circles are the VB spectral intensities integrated in the rectangular areas of panel (e). The pump fluence is 0.4mJ/cm$^2$ (corresponding to$\sim5\times10^{19}$cm$^{-3}$ photo-injected carriers, see Methods). All measurements performed at RT}
\label{f2}
\end{figure*}

Fig.\ref{f2}a shows the RT ARPES map of bulk BP along the armchair direction (see Fig.\ref{f1}), following excitation by a 30fs pulse at 1.82eV. The map, recorded at positive, i.e. after pump arrival, pump-probe delay$\sim$+2.4ps, reveals also the normally unoccupied CB. Fig.\ref{f2}b reports the differential ARPES maps obtained subtracting the photoemission spectrum acquired before pump arrival from spectra at two selected pump-probe delays. Red color represents photo-induced increase of spectral weight, while blue represents depletion. The crosses labeled 1 to 4 in Fig.\ref{f2}a mark relevant features (band extrema and band sides), and their temporal evolution is reported in Fig.\ref{f2}c. While the VB (labels 1-3) shows a prompt ($<100$fs) response to pump pulses, the CB (label 4) has a slow build-up, on the ps scale, indicating that it is {\it indirectly} populated. Photo-excited $e$ decay from higher energy levels reached by the excitation pulse towards the CB bottom via $e-e$\cite{ee} and $e-$phonon\cite{ep} scattering processes. Due to the bandgap, the enhanced intensity on the VB upper border (labels 2,3) is ascribed to photo-induced broadening and/or shift of the band edge. Such enhancement has a longer lifetime on the VB sides (label 2) compared to the VB center (label 3).

These carrier dynamics are in agreement with previous TARPES investigations on BP\cite{kimura,perfetti,grioni}. Here, we focus on two open issues: BGR and VB broadening. Fig.\ref{f2}d is a zoom-in of the VB ARPES maps at +2.4ps delay. The energy distribution curves (EDCs) are normalized to their maxima in order to enhance the band shape. Through fitting (see Methods) we quantify the band dispersion, shows as open symbols. Repeating the procedure at each pump-probe delay, we can track the evolution of the VB structure. Fig.\ref{f2}e reports the VB dispersions at three selected delays: $-0.3$ps (circles), 0ps (squares) and +2.4ps (triangles). There is a$\sim20$meV red-shift at zero pump-probe delay and a subsequent photo-induced reduction of the band curvature at positive delay, pointing at an increased $h$ effective mass. The latter is consistent with the persistent spectral enhancement at the VB sides in Fig.\ref{f2}b (label 2). However, while the parabolic fit of the VB dispersion at negative delay is good (black dash line), providing a $h$ effective mass $m_{VB}\sim0.065m_0$ ($m_0$ being the free $e$ mass) consistent with literature\cite{morita,perfetti}, the curves at zero and positive delays cannot be fitted by simple parabolas. This suggests two alternative scenarios: (i) loss of parabolic shape, particularly evident at positive delay, due to a flattening of the top VB (for $k_x<0.03$\AA$^{-1}$), rather than increase of effective mass; (ii) pump pulses induce an upward shift ($\sim50$meV) of the VB toward $E_F$, signature of BGR\cite{bgr2,bgr3}. The flattening counterbalances the BGR on the very top of the VB.
\begin{figure*}
\centerline{\includegraphics[width=180mm]{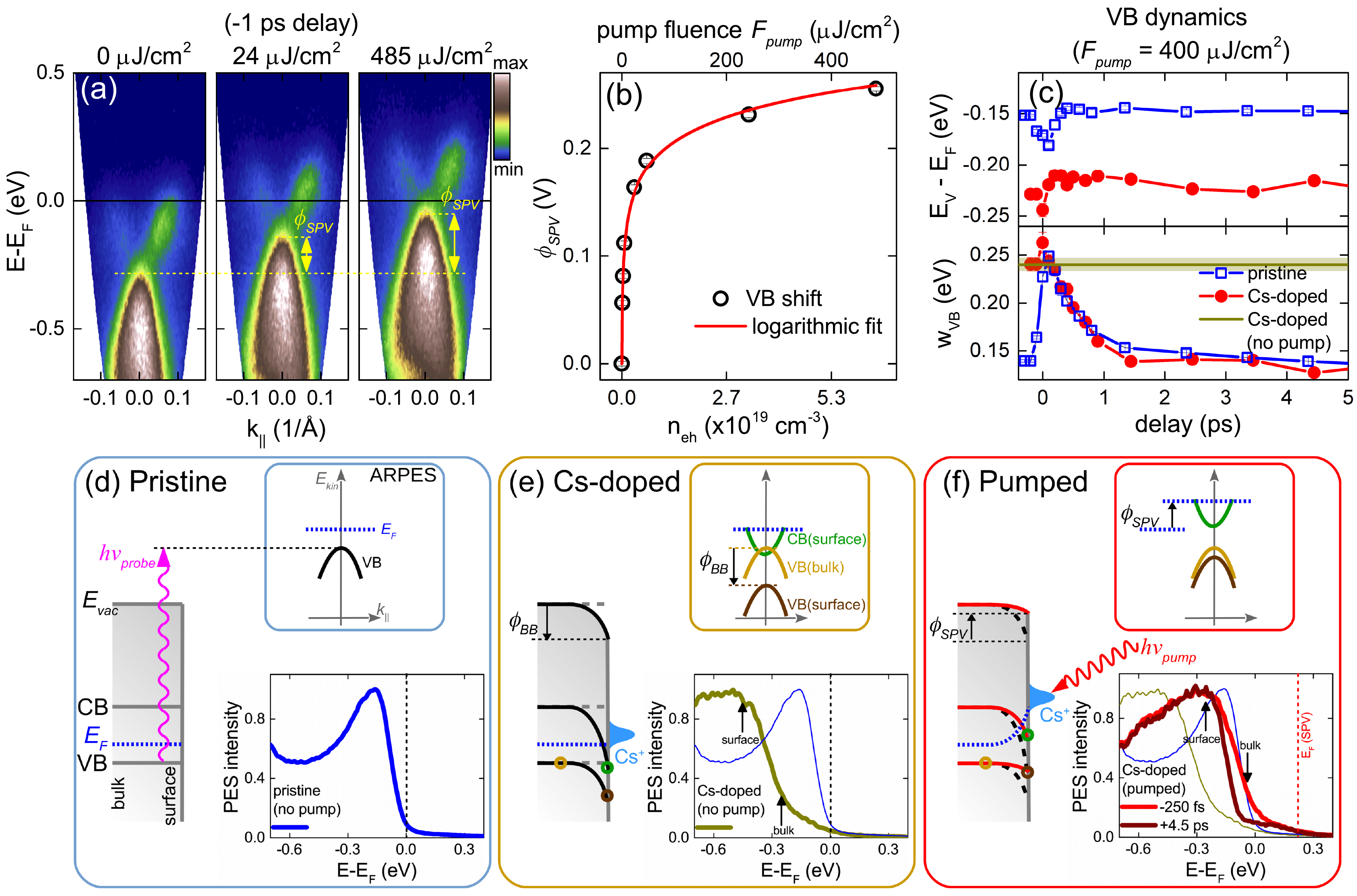}}
\caption{(a) ARPES maps after 45s Cs exposure at three pump fluences and negative pump-probe delay$\sim-1$ps. The asymmetric intensity of the CB is a matrix element effect in the photoemission process\cite{perfetti}. The shift of the spectra is the result of SPV. (b) SPV vs photo-injected carrier density (bottom scale)/pump fluence (top scale). (c) VB maximum (top) and broadening (bottom) of pristine (open squares), and biased BP (solid circles) vs delay. (d) Scheme of photoemission process for pristine BP, (e) Cs-doped BP and (f) pumped, Cs-doped BP. The upper inset in each panel sketches the expected ARPES spectrum, the lower reports the measured EDCs at $k_{||}=0$~\AA$^{-1}$. $\phi_{BB}$: band bending potential, $\phi_{SPV}$: surface photo-voltage.}
\label{f3}
\end{figure*}
\begin{figure*}
\centerline{\includegraphics[width=180mm]{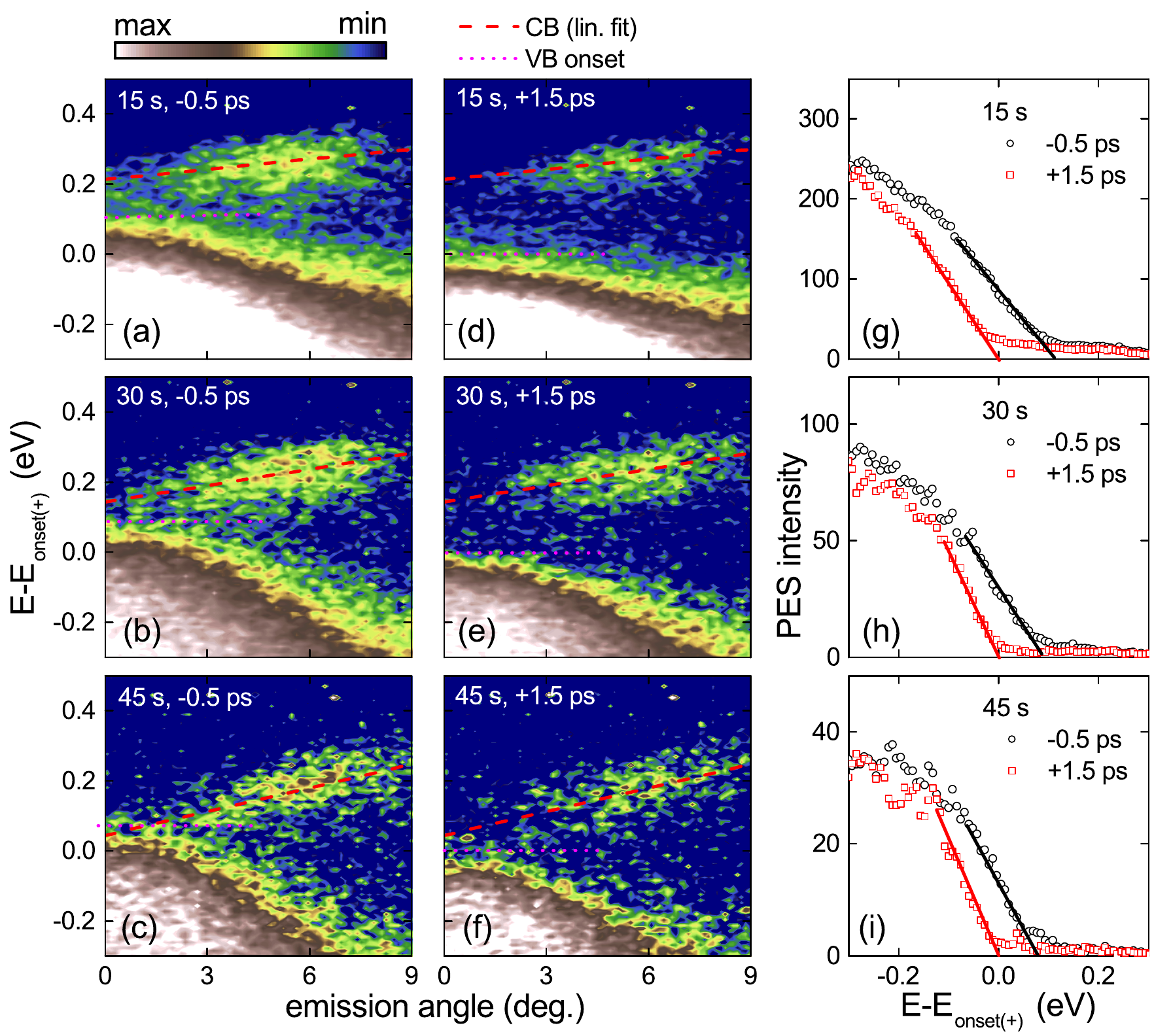}}
\caption{RT ARPES maps at (a-c) $-0.5$ps and (d-f) $+1.5$ps delay for increasing Cs exposure times (15, 30, 45s, top to bottom). (g-i) Corresponding EDCs at $0^{\circ}$ emission angle ($k_x=0$). In all maps, zero energy is referred to the VB {\it onset} at positive delay (labeled $E_{onset(+)}$). Pump fluence is$\sim$0.4mJ/cm$^2$}
\label{f5}
\end{figure*}

Further evidence is provided by the temporal evolutions in Fig.\ref{f2}f. Open symbols represent the dynamics of the VB at $k_x=0$\AA$^{-1}$, while solid squares refer to $k_x=0.044$\AA$^{-1}$ (see the corresponding dash-dotted vertical lines in Fig.\ref{f2}e). A phenomenological model based on two components, labeled C1 and C2, reproduces the observed behaviors. The positive component C1 (orange dashed line) shifts the VB towards $E_F$ and represents BGR. As photo-excited $e$ and $h$ rearrange in real space, their energies in the respective bands reduce as a consequence of screened exchange and correlation effects\cite{bgr2,bgr3,bgr1}, shrinking the bandgap. This is independent of $k_x$, with a maximum amplitude$\sim47\pm5$meV, a rise time$\sim280\pm100$fs, and a decay time$\sim4.8\pm0.8$ps.

The negative component C2 (green dotted line) has a maximum amplitude$\sim-36\pm5$meV, a pulsewidth-limited rise time ($<100$fs) and a $k_x$-dependent decay ($\sim1$ps at $k_x=0.044$\AA$^{-1}$, $\sim6$ps at $k_x=0$\AA$^{-1}$). To clarify its origin we evaluate the temporal evolution of the VB {\it intensity} by integrating the photoemission spectral weight in the small rectangular areas in Fig.\ref{f2}e. The result, displayed as green dots in Fig.\ref{f2}f, reproduces the VB transient depletion (the vertical axis on the right-hand-side of Fig.\ref{f2}f shows the relative intensity $I/I_0$, normalized to $I_0$ at negative delay). There is a one-to-one correspondence with the component C2. As photo-excited carriers ($e$ in CB and $h$ in VB) thermalize, they fill states at the respective band edges. $h$ gather at the VB maximum, as indicated by the slower intensity recovery at $k_x=0$\AA$^{-1}$. $h$ accumulation reduces the ARPES spectral weight and, consequently, the VB edge shifts downward to higher binding energy. Our pump fluence generates photo-carrier density$<10^{19}$cm$^{-3}$, largely exceeding the equilibrium carrier concentration$\sim10^{16}$cm$^{-3}$ (see Methods). The depletion of the VB top and the concurring filling of the CB bottom lead to a blue-shift of the optical absorption edge, known as Burstein-Moss shift (BMS)\cite{bm1,bm2} caused by Pauli blocking\cite{pauli}. This was observed in BP, suggesting its use as active material for mid-infrared optoelectronic devices, such as tunable infrared emitters\cite{bmbp1} and tunable optical modulators\cite{bmbp2}. BMS and BGR can have similar and opposite amplitudes for a given photo-carrier density\cite{bms}, nearly canceling each other. The coexistence of these two compensating phenomena explains the diverging conclusions on the occurrence of BGR in BP from previous TARPES investigations\cite{perfetti,grioni}. Our analysis confirms the presence of both BGR and BMS.

In order to understand the physics underlying the VB broadening, we analyze the photo-induced response of BP under the effect of a vertical electric bias, chemically induced by Cs adsorption. Since alkali atoms are $e$ donors\cite{sze}, a {\it n}-type surface region forms after doping, leading to band bending\cite{luth}. All energy levels bend downwards when moving from bulk to surface (it would be upwards for acceptor surface doping\cite{luth}). Owing to the giant Stark effect, in BP the CB bending is enhanced relative to VB\cite{liu3,kim,swkim}, eventually causing surface band inversion at sufficiently high Cs coverage ($>0.35$1L)\cite{kim}. Our TARPES analysis reveals that photoexcitation of Cs-adsorbed BP induces up to$\sim0.25$V SPV\cite{perfettiSPV,spv}. The built-in potential generated by surface doping spatially separates photo-injected $e$ and $h$. With downward band bending, $e$ migrate to the surface, while $h$ move towards the bulk, developing a dipole field (and potential) that neutralizes the alkali-induced bending. Such dipole field can extend outside the sample, accelerating photo-emitted $e$ (thus shifting all energy levels) even if they are emitted {\it before} pump arrival, provided they have not escaped the region in vacuum where the field spreads (see Methods for details).

Fig.\ref{f3}a shows three ARPES maps of BP measured after 45s Cs exposure ($\sim0.35$1L, see Methods) at {\it negative} pump-probe delay$\sim-1$ps and with increasing pump fluence. Without pump (left map) the Cs-induced modifications can be assessed by comparing with Fig.\ref{f2}a. Not only the top of the VB shifts down by$\sim0.3$eV, but also the CB minimum is now roughly {\it touching} the VB maximum, corresponding to a downward bending$\sim0.6$eV and to the apparent closure of the gap. The middle and right maps of Fig.\ref{f3}a show how increasing the pump fluence $F_{pump}$ leads to a rigid, non-linear shift of the whole ARPES spectrum to higher energy, caused by SPV. Also $E_F$ shifts accordingly. The saturation of SPV vs $F_{pump}$ is shown in Fig.\ref{f3}b. The fluence on the top axis is converted into photo-injected carrier density on the bottom axis (see Methods) and the data are fitted by a phenomenological logarithmic model\cite{widdra,marsi}: $\phi_{SPV}=\alpha k_B T/e\ln(1+n_{eh}/p_0)$. Here, $k_B T=25$~meV is the thermal energy at RT, $n_{eh}$ is the photo-generated carrier density, $\alpha=1.33\pm0.05$ and $p_0=2.8\pm0.5\times10^{16}$~cm$^{-3}$ are fitting parameters (the analog of the ideality factor in a Schottky diode\cite{marsi} and the equilibrium majority carrier density, respectively).

Transient photo-induced effects are reported in Fig.\ref{f3}c, where we compare the VB dynamics, i.e. the energy of its maximum $E_V$ and the edge width $w_{VB}$ deduced from the fitting procedure, before (open squares) and after (solid circles) Cs exposure, for the same excitation fluence$\sim0.4$~mJ/cm$^2$. The evolution of $E_V$ for pristine BP is the same as Fig.\ref{f2}f. Apart from the different binding energies, the photo-induced dynamics of the VB maximum (upper graph) before and after Cs doping are very similar. Thus, BGR and BMS appear to coexist in the biased sample. More challenging to understand is the dynamics of the edge width (lower graph). Photo-injection in pristine BP leads to a prompt VB broadening ($w_{VB}$ nearly doubles within$\sim$150fs), as for Ref.\citenum{perfetti}. In biased BP, before optical excitation, the VB width is almost twice the pristine case (dark yellow horizontal line in Fig.\ref{f3}c), while after pumping (solid circles) it drops, following the dynamics in the unbiased sample.

These behaviors can be understood with the help of Figs.\ref{f3}d-f that sketch the photoemission process under specific situations. Fig.\ref{f3}d depicts the pristine BP case. All energy levels (CB, VB, $E_F$ and vacuum level $E_{vac}$) are represented as horizontal lines since no band bending is present. Probe photons $h\nu_{probe}$ promote bound $e$ from occupied VB states to free $e$ that can travel in vacuum towards the analyzer where the ARPES map is recorded, as sketched in the upper inset. The measured EDC at $k_x=0$\AA$^{-1}$ is shown in the lower inset.

Fig.\ref{f3}e depicts the situation after alkali adsorption. Cs surface states are ionized donors and therefore lie above $E_F$. As a consequence, the $n$-doped surface region of the sample is characterized by the downward bending potential $\phi_{BB}$. Owing to Stark effect, the CB bending is enhanced with respect to the VB one. At sufficient doping ($>0.1$1L according to Ref.\citenum{kim}), the minimum of the CB at the surface falls below $E_F$ (green circle) and can be observed in static photoemission. Focusing on the VB, $e$ photoemitted from the surface (brown circle) have the largest binding energy and determine the main peak of photoemission (see EDC in the lower inset of Fig.\ref{f3}e). However, with the 6eV photons used here, bulk sensitivity is enhanced due to the large $e$ mean free path ($\sim10$nm)\cite{inelastic}. Accordingly, bulk VB states (dark yellow circle) contribute to the spectral weight with lower binding energy. The enhanced VB width after Cs doping is caused by these bulk states, see upper inset of Fig\ref{f3}e, marked by the vertical arrows in the EDC.

Fig.\ref{f3}f shows the pump effect on Cs-doped BP. As SPV develops, all energy levels at the surface shift upward, compensating the bending potential. Here, a distinction between negative and positive pump-probe delays must be made. If probe photons {\it precede} the pump pulse, $e$ are emitted before SPV sets in. A few ps after pumping, the photovoltage creates a dipole field that extends in vacuum, accelerating the traveling free $e$ and rigidly shifting the photoemission spectrum. Within the experimental uncertainty, the corresponding EDC is identical, apart from the energy shift, to that recorded without pump, since $e$ are emitted before excitation, thus preserving the un-pumped spectral shape (as can be confirmed by comparing the red and dark yellow EDCs in the lower inset). Instead, at positive pump-probe delay SPV develops {\it before} photoemission, surface and bulk states level to similar binding energies (see upper inset) and the photoemission spectrum resembles the pristine BP case with sharper VB edge, as experimentally observed in the dark red EDC in the lower inset. Based on these considerations, the dynamics of the VB width in Fig.\ref{f3}c (red curve) represents the SPV temporal onset which, in turn, reflects the dynamics of $e-h$ space separation that builds up the dipole field\cite{sobota}.

Having clarified the physical origin of the observed spectral features, we can now explain the ubiquitous photo-induced VB broadening. The close resemblance of the VB edge temporal evolution in pristine and Cs-doped samples at positive delay, Fig.\ref{f3}c (bottom), points at a common mechanism, regardless of bias. Pump pulses produce identical $e$ and $h$ distributions that can largely exceed the equilibrium densities (see Methods) and decay exponentially with depth. Density gradients trigger carriers diffusion. In bulk BP, the mobility of $h$ in the direction normal to the surface is larger than $e$\cite{morita}. Since mobility is proportional to the diffusion coefficient via Einstein's relation\cite{sze}, $h$ diffuse faster than $e$. Thus, immediately after pumping, $h$ move to the bulk, leaving $e$ at the sample surface even in the absence of a vertical bias. As $e$ and $h$ spatially separate, an outward-bound dipole field develops producing a potential, called Dember photovoltage\cite{spv}, that can alter band binding energies at the surface. The build-up of this photovoltage is as fast as charge separation, and it can happen on a sub$-100$fs timescale (see Methods). Due to the bulk sensitivity of our photoemission setup, we detect these alterations as band broadening. The orientation of the dipole field opposes the diffusion process. Eventually, charge separation (thus spectral broadening) will stop, as for Fig.\ref{f3}c. Further support to this explanation is provided by two experimental observations. (i) The Dember photovoltage saturates logarithmically with photo-carrier density\cite{spv} (see Methods) and (ii) VB broadening in pristine BP shows the same logarithmic behavior with respect to pump fluence\cite{perfetti}.

Thus, the occurrence of band inversion must be cautiously claimed in biased BP. Figs.\ref{f5}a-f show ARPES maps recorded at negative ($-0.5$ps, a-c) and positive ($+1.5$ps, d-f) pump-probe delays for increasing Cs exposure times (15, 30, 45s, top to bottom). The corresponding EDCs at $k_x=0$\AA$^{-1}$ are in Figs.\ref{f5}g-i. To harmonize the comparison, the zero binding energy is referred to the {\it onset} of the VB at positive delay (labeled $E_{onset(+)}$). Owing to Stark effect\cite{kim,swkim}, as Cs concentration increases, the CB minimum downshifts and falls within the onset of the VB at negative delay for the highest doping, reached at 45s (Fig.\ref{f5}c). This corresponds to the critical Cs coverage$\sim$0.35 1L identified in Ref.\citenum{kim} at which VB and CB cross. Comparing negative and positive delays, we observe that, after photo excitation, the VB downshifts, seemingly leading to a larger bandgap. Recalling Figs.\ref{f3}e-f, this is the effect of SPV that compensates the Cs-induced bending potential. The {\it apparent} VB shift is due to its edge sharpening. Before pumping, the VB onset is determined by bulk states having smaller (less negative) binding energy as compared to surface states (see upper inset of Fig.\ref{f3}e). After excitation, SPV neutralizes band bending, thus bulk states "align" to surface states, appearing at larger binding energy (see upper inset of Fig.\ref{f3}f) and resulting in a sharper VB edge. The crossing of VB and CB at $-0.5$ps with the largest Cs dose (Fig.\ref{f5}c) is the overlap of {\it bulk} VB and {\it surface} CB, resolved thanks to the bulk sensitivity of our ARPES photon energy. The CB binding energy and line shape do not significantly vary upon pumping. We only observe a loss of spectral weight after optical excitation, due to photo-induced depletion. Considering that CB bending is more pronounced than VB, due to the Stark effect, and that SPV affects all bands equally, we should expect a well discernible photo-induced sharpening of CB at positive delay, which is not observed. The width invariance of the CB vs delay confirms its strong surface confinement, in agreement with Refs.\citenum{kim,ehlen,swkim,hof}. Thus, our data demonstrate that band crossing in chemically biased bulk BP should be claimed with caution: only in 1L or FL-BP ($\leq4$ layers\cite{swkim}) band inversion could be seen, since bulk contributions are absent.
\begin{figure*}
\centerline{\includegraphics[width=180mm]{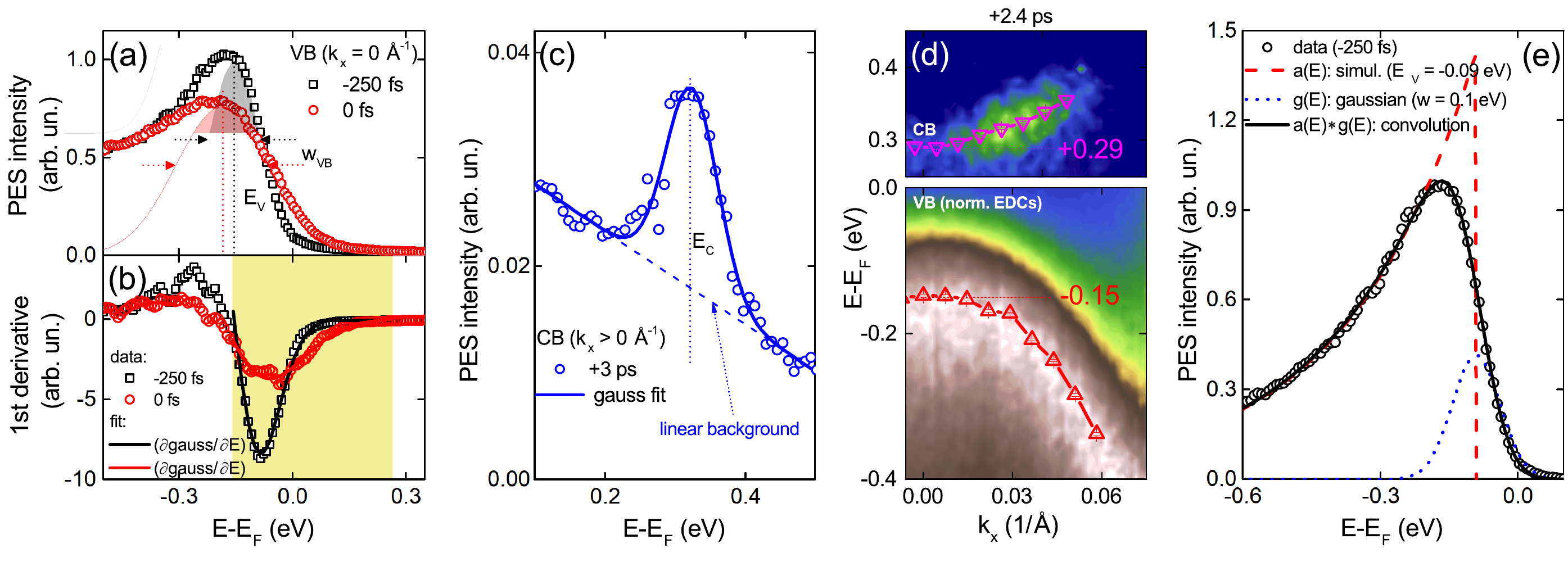}}
\caption{(a) VB EDCs of pristine BP at $k_x=0$~\AA$^{-1}$ for $-250$fs (black) and 0fs (red) pump-probe delays. (b) Numerical derivative of EDCs shown in (a): the solid lines are fits according to Eq.\ref{gauss}. The resulting peaks are reproduced as shaded areas in panel (a), with the corresponding peak positions $E_V$ and peak widths $w_{VB}$. (c) EDC of the CB at $k_x=0.04$\AA$^{-1}$ and $+3$ps delay. The solid line is a gaussian fit centered at $E_C$. (d) ARPES maps and dispersion fits of CB (top) and VB (bottom) at 2.4ps delay. Band extrema are indicated. (e) Simulation of VB photoemission spectrum at negative delay. The spectral function $a(E)$ is peaked at $E_V-E_F=-0.09$eV and decays exponentially at larger binding energy. Once convoluted with a gaussian profile $g(E)$ (line width $w=0.1$eV), it matches experiments.}
\label{si1}
\end{figure*}
\section{Conclusions}
We employed time- and angle-resolved photoemission spectroscopy to investigate the transient electronic dynamics of bulk BP. Our measurements show photo-induced bandgap renormalization entwined to VB flattening, caused by Pauli blocking. Applying a vertical electric bias, chemically induced by Cs ad-sorption, we showed that the ubiquitous VB broadening is due to photo-carriers ambipolar drift-diffusion. Both VB and (to a larger extent) CB experience surface bending upon doping, that can be counterbalanced by a surface photovoltage prompted by laser irradiation. This allows us to unequivocally discriminate bulk vs surface states, to establish the surface localization of the conduction band and to elucidate the occurrence of band inversion in bulk samples. For any application of black phosphorus involving hot carrier injection, transient changes will affect device performance. Our analysis reveals a rich and multifaceted photo-induced band dynamics that might help design opto-electronic devices. Since broad gap tunability by alkali atoms ad-soprtion has been demonstrated in transition-metal dichalcogenides\cite{tmd,tmd2}, our results will be relevant to a wider class of layered materials.
\section{Acknowledgements}
We acknowledge funding from EU Graphene Flagship, ERC Grants Hetero2D, GSYNCOR, EPSRC Grants EP/K01711X/1, EP/K017144/1, EP/N010345/1, EP/L016087/1, and Italian PRIN project 2017BZPKSZ.
\section{Methods}
\section{Spectral analysis and fitting}\label{fitting}
To quantify band dispersions and photo-induced effects, we employ a fitting routine of VB and CB measured with TARPES.

Fig.\ref{si1}a plots the EDCs of the VB at $k_x=0$\AA$^{-1}$ for $-250$fs (black) and 0fs (red) pump-probe delays. Photo-induced depletion and shift/broadening of the VB are seen. The peak asymmetry, caused by secondary $e$ (inelastic scattering events in the photoemission process\cite{luth}), makes the exact peak identification ambiguous. Therefore, we employ a different approach. We first compute the EDCs numerical derivative, then we fit the resulting curves with the analytical derivative of a gaussian profile:
\begin{equation}
\frac{\partial}{\partial E}\left[Ae^{-\frac{2(E-E_V)^2}{w_{VB}^2}}\right]=-\left[\frac{4A (E-E_V)}{w_{VB}^2}\right]e^{-\frac{2(E-E_V)^2}{w_{VB}^2}} \label{gauss}
\end{equation}
where $E_V$ is the binding energy, $w_{VB}$ is the width ($2\sigma$) and $A$ is the peak amplitude. Fig.\ref{si1}b reports the results. The fits in the yellow-shaded region provide an accurate determination of the high-energy side of the original VB peaks, as testified by the reconstructed gaussian profiles in Fig.\ref{si1}a (colored shaded areas), and overcome the peak asymmetry issue. To deduce the VB dispersion, this procedure is repeated for various $k_x$. Using the Shirley method\cite{shirley} to remove the incoherent background leads to very similar peak positions and widths, but poorer estimates of peak amplitude. Fig.\ref{si1}c reports the fitting routine for the CB (at +3ps delay and $k_x \neq 0$~\AA$^{-1}$, where it is more evident). In this case, the peak is modeled by a gaussian profile with a linear background, since the spectral feature is well-defined. The CB binding energy is given by the position of the gaussian peak ($E_C$). Similarly to VB, the CB dispersion is obtained repeating the fit at various $k_x$.
\section{Electronic and optical properties of BP}\label{properties}
Using the fitting routine previously described, we estimate the BP bandgap $E_g=0.44\pm0.01$eV, with VB maximum$\sim0.15$eV below $E_F$ and CB minimum$\sim0.29$eV above $E_F$, see Fig.\ref{si1}d. Data refer to positive pump-probe delay$\sim$2.4ps when CB is populated. This is the (VB)peak-to-(CB)peak energy gap. Refs.\citenum{grioni,onset1,onset2} suggest to use the (VB)onset-to-(CB)onset as definition of bandgap. This would give $E_g\sim0.3$eV. However, considering the combined energy-time resolution of our experiments, the use of band onsets underestimates the gap. On the other hand, $E_g=0.44\pm0.01$eV is slightly larger than the commonly reported values ($0.3-0.4$eV)\cite{liu}. The discrepancy might be caused by the $k_z$ sensitivity of photoemission, related to our photon energy. Owing to the strong band dispersion along the $\Gamma Z$ crystallographic direction (see Fig.4c), our 6eV probe photon might correspond to a $k_z$ slightly away from the $Z$-point, thus detecting a larger bandgap. We also show a VB spectral simulation assuming a gaussian line width $w\sim0.1$eV, smaller than the measured one ($w_{VB}\sim0.15$eV) convoluted with a possible spectral function $a(E)$, Fig.\ref{si1}e. $a(E)$ decays exponentially at large binding energy and peaks at $E_V-E_F=-0.09$eV. Once convoluted with a gaussian profile, it provides an excellent fit of the experimental spectrum, placing the VB maximum at $E_V-E_F=-0.09$eV instead of $-0.15$eV, with $E_g=0.38$eV.

Regardless of the exact value of $E_g$, equilibrium carrier concentrations in BP at RT can be estimated\cite{am}: $n_0=N_Ce^{-(E_C-E_F)/k_BT}$ ($e$ in CB), $p_0=N_V e^{-(E_F-E_V)/k_BT}$ ($h$ in VB) and $n_i^2= n_0 p_0=N_C N_{V}e^{-(E_C-E_V)/k_BT}$, with $N_{C}=2.5 (m_C/m_0)^{3/2}\times10^{19}$cm$^{-3}$, $N_{V}=2.5(m_V/m_0)^{3/2}\times 10^{19}$cm$^{-3}$. Here, $E_V$ and $E_C$ are the energies of the VB maximum and CB minimum, $m_V$ and $m_C$ the respective effective masses and $n_i$ is the intrinsic carrier density. Due to the very similar $e$ and $h$ effective masses in BP ($m_C\sim m_V\sim0.23m_0$)\cite{morita}, we get $N_C\sim N_V\sim 2.8\times10^{18}$cm$^{-3}$. Using the measured bandgap and VB maximum, we estimate the equilibrium $e$ and $h$ densities (for comparison, we use values obtained from the fitting procedure and from the convolution example of Fig.\ref{si1}e) in Table\ref{t1}.
\begin{table}[h!]
\begin{tabular}{c|c|c|c|c}
$E_g$[eV] & $E_V-E_F$[eV] & $n_{i}$[cm$^{-3}$] & $p_0$[cm$^{-3}$] & $n_0$[cm$^{-3}$] \\
\hline
$0.44\pm0.01$ & -0.15 & $4.2\times10^{14}$ & $6.9\times10^{15}$ & $2.6\times10^{13}$ \\
$0.38$ & -0.09 & $1.4\times10^{15}$ & $7.7\times10^{16}$ & $2.6\times10^{12}$ \\
\end{tabular}
\caption{$E_g$ bandgap, $E_V-E_F$ VB binding energy, $n_{i}$ intrinsic carrier density, $p_0$ and $n_0$ estimated $h$ (in VB) and $e$ (in CB) densities at RT.}
\label{t1}
\end{table}

\begin{figure}
\centerline{\includegraphics[width=90mm]{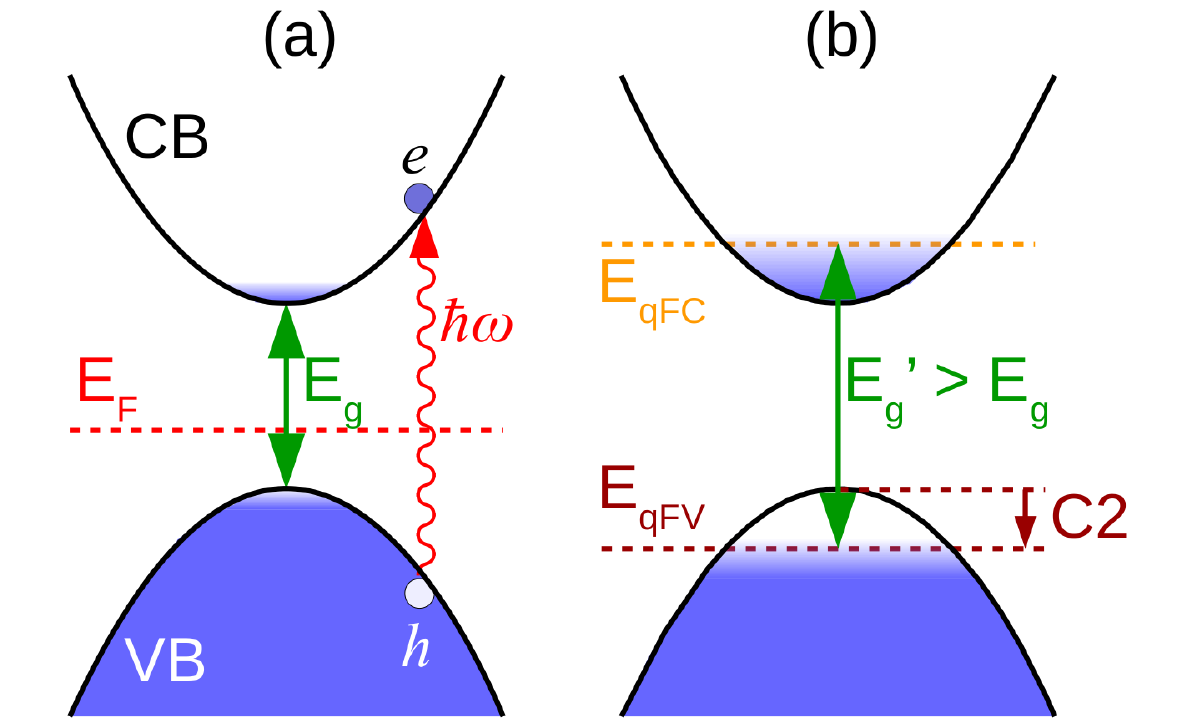}}
\caption{Band filling effect. (a) In equilibrium VB is fully occupied and CB empty. Photo-excitation promotes $e$ to CB leaving $h$ in VB. (b) After intraband thermalization, the VB maximum is depleted (i.e. filled by $h$) and the CB minimum is occupied, leading to a larger bandgap $E'_g$ defined by the corresponding quasi-Fermi levels}
\label{bm}
\end{figure}

The majority carrier density $p_0\sim10^{16}$~cm$^{-3}$ agrees with that obtained from the fluence dependence of SPV (Fig.2b). It is instructive to compare this value with the photo-induced $e-h$ density. According to the optical properties of BP\cite{morita}, with pump photon $h\nu\sim1.82$eV ($\lambda\sim680$nm) and electric field polarized along the armchair direction, the dielectric constant is $\varepsilon\sim12+2i$, which leads to refractive index $n=\sqrt{\varepsilon}\sim3.5+0.3i$, reflectivity $R=(|n-1|/|n+1|)^2\sim0.3$ and absorption length $1/\alpha=\lambda/4\pi$Im$(n)\sim180$nm\cite{vab}. The incident pump fluence $F_{pump}$[J/cm$^2$] can be converted into density of photo-generated $e-h$ pairs $n_{eh}$[cm$^{-3}$] using the relation\cite{vab} $n_{eh}=F_{pump} (1-R)\alpha/h\nu$ (assuming quantum efficiency of 1). Similarly, the increase of the lattice temperature can be estimated as\cite{vab}: $\Delta T=F_{pump} (1-R)\alpha M/\rho c_p$, where $\rho\sim2.7$g/cm$^3$\cite{thermal} is the mass density, $M\simeq31$g/mol is the molar mass and $c_p\simeq 21$J/mol K is the BP lattice specific heat\cite{thermal}. With pump fluence$\sim0.5$mJ/cm$^2$ we obtain $n_{eh}\sim7\times10^{19}$cm$^{-3}$ and $\Delta T\sim11$K. Although the lattice temperature is hardly affected by the laser irradiation, $n_{eh}$ (photo excitation)$\gg p_{0}$ (majority carrier density). Therefore, with our laser fluence we are always in a strong electronic photo-excitation regime, and state filling effects cannot be ignored. At equilibrium, the VB is almost fully occupied, and the CB empty (apart from the mild thermal carrier populations), Fig.\ref{bm}a. Illumination promotes $e$ to CB, leaving $h$ in VB. As carries thermalize, they occupy the respective band edges, Fig.\ref{bm}b.
\begin{figure}
\centerline{\includegraphics[width=90mm]{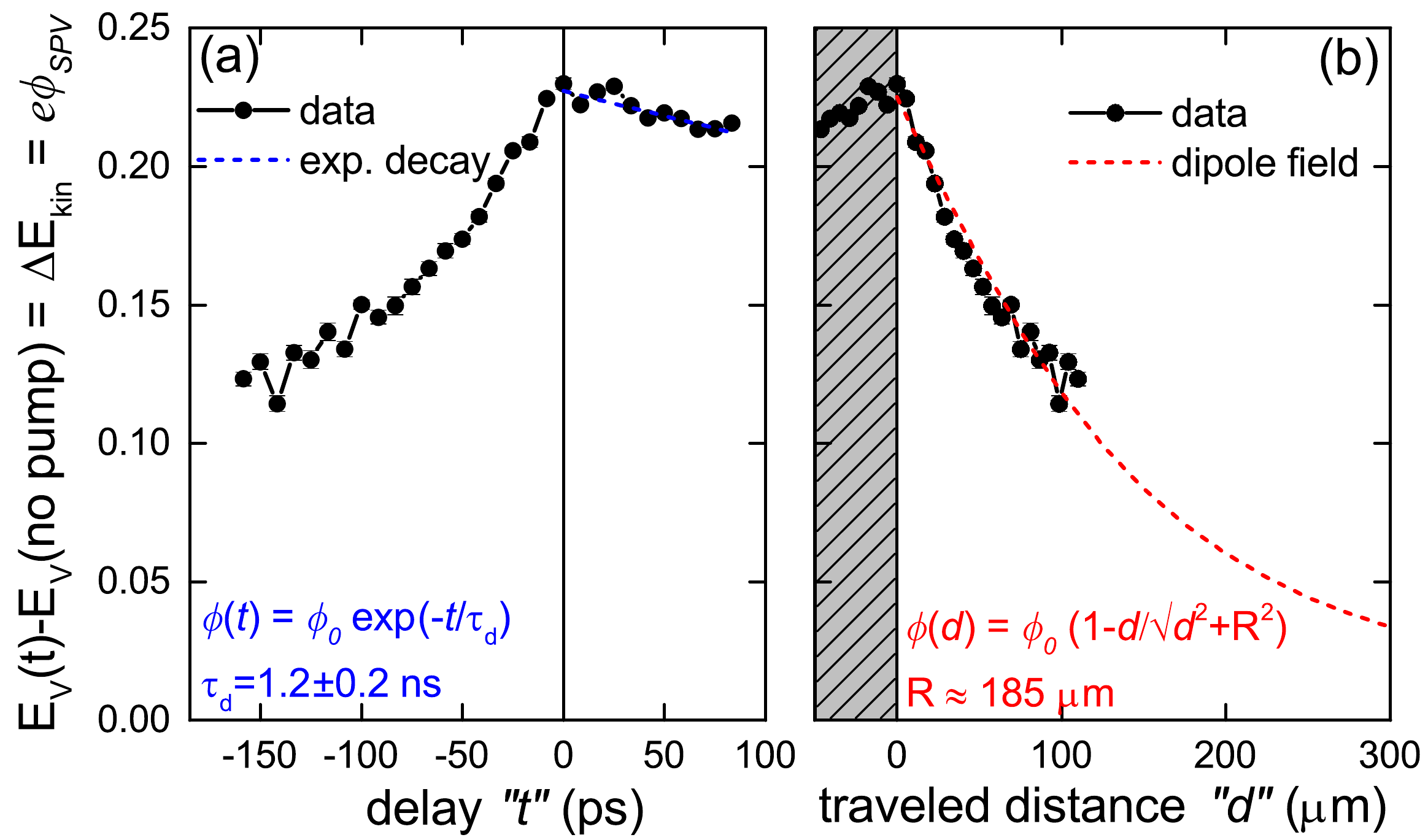}}
\caption{Measured PES kinetic energy of VB as a function of (a) pump-probe delay and (b) distance of the photo-emitted $e$ from sample surface. Dash line is the model from Eq.\ref{tanaka}}
\label{si2}
\end{figure}
\begin{figure*}
\centerline{\includegraphics[width=180mm]{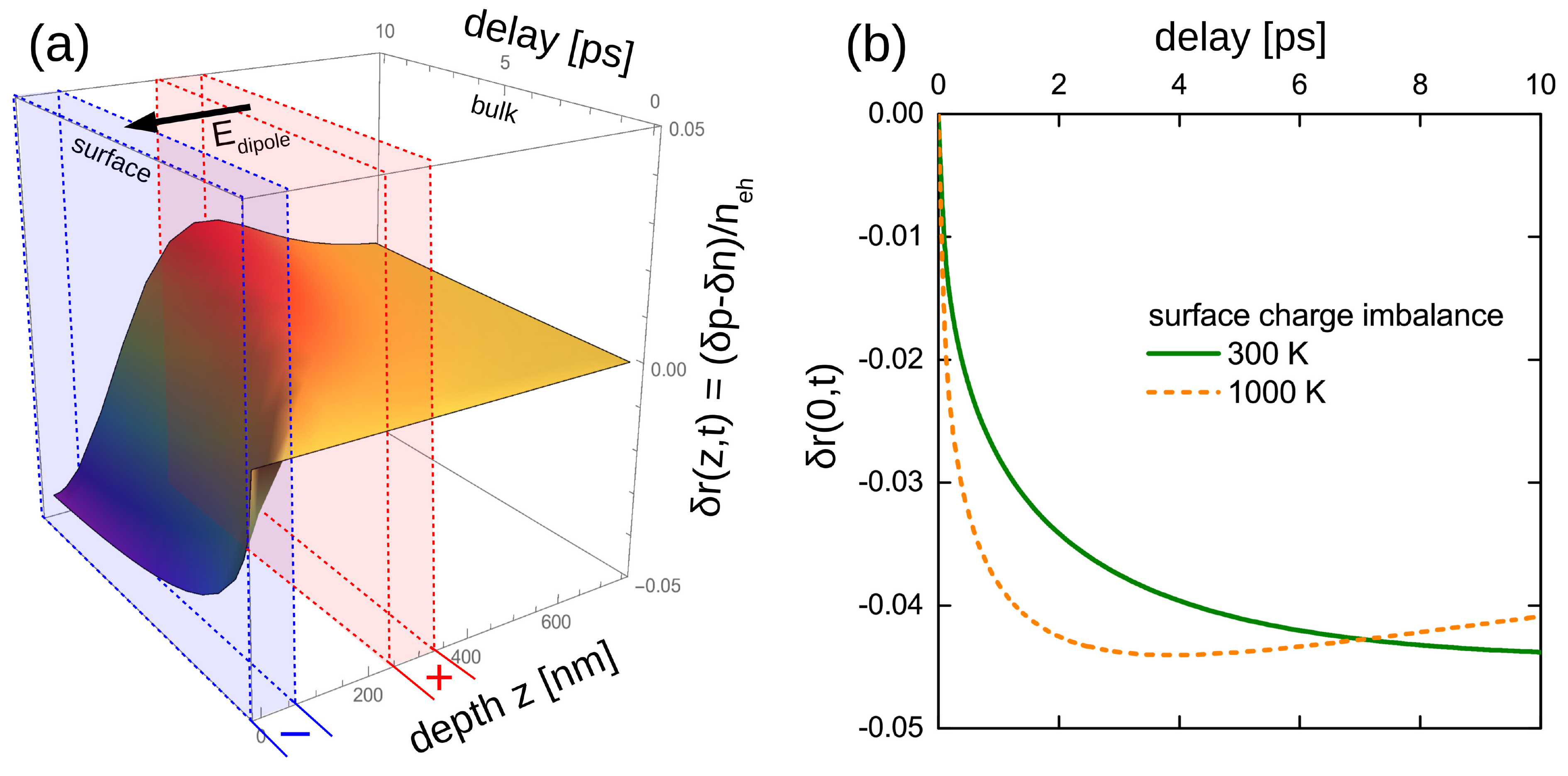}}
\caption{(a) Simulation of $e/h$ diffusion after optical excitation, according to Eqs.\ref{d1}-\ref{bc}. The colored surface represents $[\delta p(z,t)-\delta n(z,t)]/n_{eh}$ as a function of delay $t$ and depth $z$ (blue color: $p<n$, red color $p>n$). Charge separation induces a dipole field $E_d$ that counterbalances diffusion. (b) Surface charge imbalance at $z=0$ as a function of delay at 300K (green-solid) and 1000K (orange-dashed)}
\label{si3}
\end{figure*}

The resulting out-of-equilibrium charge distributions can be defined by the so-called {\it quasi}-Fermi levels for CB ($E_{qFC}$) and VB ($E_{qFV}$)\cite{spv,shur}, both departing from the equilibrium Fermi energy $E_F$\cite{bm5}. State occupancy increases the optical bandgap to $E_g'=E_{qFC}-E_{qFV}$ (Burstein-Moss effect\cite{bm1,bm2} caused by Pauli blocking\cite{pauli2}) pushing down (up) the VB (CB) edge. This explains the origin of component C2 in Fig.1f.
\section{Surface photovoltage (SPV)}\label{surfacepv}
SPV has been extensively investigated since the early 1950s\cite{brat1,brat2,brat3,spv1}. It is ubiquitous in doped semiconductors\cite{monch}. The basic ingredient is the presence of (intrinsic or doping-induced) surface/interface states acting as $e$ donors or acceptors. The consequence is the formation of a space-charge layer. The resulting band bending is estimated solving Poisson's equation $\nabla^2 \phi_{BB} \propto p-n+N_D^+-N_A^-$ ($\phi_{BB}$ is the built-in potential, $p$, $n$, $N_D^+$ and $N_A^-$ are $h$, $e$, donor and acceptor densities, respectively)\cite{monch}. Light absorption with photon energy larger than the gap generates $e-h$ pairs and the electric field in the space-charge layer spatially separates opposite charges, leading to band flattening. The band bending is {\it compensated} by the photo-injection. This is the SPV effect. Following Ref.\citenum{spv}, several approaches have been developed. Here, we point out some aspects involving time-resolved photoemission techniques. The logarithmic dependence of SPV with photo-injected carrier density was suggested in Ref.\citenum{spv1}, and adapted to time-resolved experiments in Refs.\citenum{widdra,marsi,tr2}. Although some constraints apply when dealing with time-resolved photoemission\cite{tanaka}, the technique can provide useful information. (i) The SPV measured at negative pump-probe delay allows one to retrieve the spatial dependence of the electric field (and potential) outside the sample surface\cite{tanaka}. (ii) The band energy shift measured at positive delay allows one to deduce the SPV lifetime\cite{sobota}. At negative delay $t<0$, $e$ are emitted {\it before} pump arrival. They travel in vacuum with kinetic energy $m_0 v^2/2$ (and velocity $v$) covering a distance $d=|vt|$ from the sample surface. At that point, the pump pulse reaches the sample and triggers SPV. The resulting dipole field (and dipole potential) extends in vacuum, accelerating the traveling electrons. The gain in kinetic energy $\Delta E_{kin}$ is proportional to the intensity of dipole potential $\phi$ at distance $d$, i.e. $\Delta E_{kin}(d)=e\phi(d)$. Therefore, mapping $\Delta E_{kin}$ vs distance provides the spatial profile of the dipole potential in front of the surface.

Fig.\ref{si2}a depicts the change of VB binding energy (relative to that measured {\it without} pump, i.e. with no SPV) in Cs-doped BP vs delay. It embodies the gained kinetic energy as a function of delay. At $t>0$ it mildly decays and an exponential fit provides the lifetime of SPV $\tau_d=1.2\pm0.2$ns (the temporal measurement window is limited by the travel range of our delay stage). Fig.\ref{si2}b reports the same data plotted as a function of traveled distance $d=|v t|$ for $t<0$, and $v$ deduced from the measured kinetic energy of the VB {\it without} pump. The dashed red line at $d>0$ is the fit with the electrostatic potential $\phi$ generated by a uniform dipole distribution on a disc of radius $R$\cite{tanaka}:
\begin{equation}
\phi(d)=\phi_0\left(1-\frac{d}{\sqrt{d^2+R^2}}\right)\label{tanaka}
\end{equation}
The resulting disc diameter $2R$ is$\sim3$ times larger that the pump spot size. This is due to the non-linear saturating behavior of SPV with light intensity. Even at the border of the laser spot, where laser intensity is weaker, the SPV might be as large as in the spot center, resulting in an {\it apparent} disc diameter larger than the nominal laser spot size.
\section{VB broadening in pristine BP}\label{broadening}
The pump pulse produces identical $e$ ($\delta n$) and $h$ ($\delta p$) distributions that can largely exceed the equilibrium densities ($p_0,n_0$), especially at the surface (i.e. $\delta p,\delta n\gg p_0\gg n_0$ for $p$-doped samples), and decay exponentially with depth. The density gradient triggers carriers diffusion. For simplicity, here we will ignore recombination and drift. According to Ref.\citenum{morita}, in bulk BP $h$ mobility ($\mu_p$) along the $z$-axis, normal to the surface, is higher than $e$ ($\mu_n$). Since mobility is proportional to the diffusion coefficient\cite{sze}, $h$ diffuse faster than $e$. Thus, after pumping, $e$ tend to accumulate at the sample surface, while $h$ move to the bulk. This can be simulated via diffusion equation\cite{mac}:
\begin{eqnarray}
\frac{\partial \delta n(z,t)}{\partial t} &=& D_n \frac{\partial^2 \delta n(z,t)}{\partial z^2} \label{d1} \\
\frac{\partial \delta p(z,t)}{\partial t} &=& D_p \frac{\partial^2 \delta p(z,t)}{\partial z^2} \label{d2} \\
\delta p(z,0) &=& \delta n(z,0) = n_{eh} \exp(-\alpha |z|) \label{bc}
\end{eqnarray}
Where $D_n=k_B T\mu_n/e$, $D_p=k_BT\mu_p/e$; $n_{eh}$ and $\alpha$ are the photo-excited carrier density and the pump absorption coefficient, respectively. The use of the absolute value $|z|$ in Eq.\ref{bc} ensures no diffusion through the surface. Eqs.\ref{d1}-\ref{d2}, with initial condition given by Eq.\ref{bc}, can be analytically solved to obtain the carrier distributions over time and depth. $\delta r(z,t)=[\delta p(z,t)-\delta n(z,t)]/n_{eh}$ is relevant for us. $\delta r>0$ indicates $h$ excess, while $\delta r<0$ represents $e$ excess. The temporal and depth evolution of $\delta r(z,t)$ are reported in Fig.\ref{si3}a.

$e$ accumulate at the sample surface within a few ps (blue color, $\delta r<0$), while $h$ move deeper in the bulk (red color, $\delta r>0$). The estimated diffusion coefficients refer to RT. Laser pumping can induce a much higher {\it electronic} T\cite{carpene}, increasing the diffusion coefficient and speeding up $e$ accumulation at the surface. Fig.\ref{si3}b shows the surface charge ratio $\delta r$ at $z=0$ as a function of delay for 300K (blue) and 1000K (red). As $e$ and $h$ separate, a dipole field develops. The area marked in light blue in Fig.\ref{si3}a represents the surface region of the sample with predominantly negative charge, while the light red area refers to the region where positive charge prevails. The electric dipole field $E_d$, marked by the large black arrow, arises between these two regions and counterbalances charge separation. We can estimate the dipole field from current balance\cite{mac}:
\begin{eqnarray}
J_n(z,t)/e &=& \mu_n (\delta n+n_0) E_d + D_n \frac{\partial \delta n(z,t)}{\partial z} \label{j1} \\
J_p(z,t)/e &=& \mu_p (\delta p+p_0) E_d - D_p \frac{\partial \delta p(z,t)}{\partial z} \label{j2}
\end{eqnarray}
The field $E_d$ that neutralizes diffusion corresponds to an overall vanishing current, i.e. $J_n+J_p=0$, leading to:
\begin{equation}
E_d = \frac{D_p \frac{\partial \delta p}{\partial z}-D_n \frac{\partial \delta n}{\partial z}}{\mu_p (\delta p+p_0)+\mu_n (\delta n+n_0)} \approx \frac{(D_p -D_n) \frac{\partial \delta p}{\partial z}}{(\mu_p + \mu_n) \delta p+\mu_p p_0} \label{dember}
\end{equation}
In the last term of Eq.\ref{dember} we use the fact that $\delta p\sim\delta n\gg p_0\gg n_0$ for a $p$-doped sample. When moving from bulk to surface, the dipole field $E_d$ builds up a potential $\phi_D$, called Dember photovoltage\cite{spv}:
\begin{eqnarray}
\phi_D &=& -\int_0^\infty E_d dz = -\int_0^\infty \frac{(D_p -D_n) \frac{\partial \delta p}{\partial z}}{(\mu_p + \mu_n) \delta p+\mu_p p_0} dz \\
         &=&\frac{D_p-D_n}{\mu_p+\mu_n} \ln\left(1+\frac{\mu_p+\mu_n}{\mu_p} \frac{\delta p(0,t)}{p_0}\right) \label{satur}
\end{eqnarray}
With $\mu_p=550$cm$^2/$Vs\cite{morita}, $\mu_n=400$cm$^2/$Vs\cite{morita} and $\delta p/p_0\sim10^2\div10^3$, we obtain $\phi_D\sim(0.8\div1.2) k_B T/e$. A transient photo-induced electronic T$\sim10^3$K would result in $\phi_D\sim0.1$V, compatible with the measured VB broadening. According to Eq.\ref{satur}, $\phi_D$ also saturates logarithmically with photo-injected carrier density, in agreement with Ref.\citenum{perfetti}. Thus, the photo-induced dynamics of the VB peak width in pristine BP can be ascribed to transient charge separation, in analogy with the case of doped BP. This explains the dynamics in Fig.2c.

\end{document}